# Observation of squeezed Chern insulator in an acoustic fractal lattice

Junkai Li[1#], Yeyang Sun[1#], Qingyang Mo[1], Zhichao Ruan[1,2*] and Zhaoju Yang[1*]

[1]Interdisciplinary Center for Quantum Information, Zhejiang Province Key Laboratory of Quantum Technology and Device,

Department of Physics, Zhejiang University, Hangzhou 310027, Zhejiang Province, China

[2]State Key Laboratory of Modern Optical Instrumentation, Zhejiang University, Hangzhou 310027, Zhejiang Province, China

#These authors contributed equally to this work

*Email: zhichao@zju.edu.cn; zhaojuyang@zju.edu.cn.

**Abstract**

Topological insulators are a new phase of matter with the distinctive characteristics of an insulating bulk and conducting edge states. Recent theories indicate there even exist topological edge states in the fractal-dimensional lattices, which are fundamentally different from the current studies that rely on the integer dimensions. Here, we propose and experimentally demonstrate the squeezed Chern insulator in a fractal-dimensional acoustic lattice. First, through calculating the topological invariant of our topological fractal system, we find the topological phase diagram is squeezed by about 0.54 times, compared with that of the original Haldane model. Then by introducing synthetic gauge flux into an acoustic fractal lattice, we experimentally observe the one-way edge states that are protected by a robust mobility gap within the squeezed topological regimes. Our work demonstrates the first example of acoustic topological fractal insulators and provides new directions for the advanced control of sound waves.

Featuring the insulating bulk and conducting edges, topological insulators[1,2] reveal a topologically distinct phase of matter that cannot be described by Landau's theory of phase transitions. One important class of topological systems is Chern insulator or quantum anomalous Hall effect, which has integer Hall conductivity and no Landau levels. In 1988, Haldane presented a paradigmatic example[3] of Chern insulator on a honeycomb lattice with staggered magnetic flux but a net magnetic field of zero. This topological Haldane model utilizes both the time-reversal and inversion symmetry breaking to open a bandgap at the Dirac points. The competition between the two broken symmetries results in topologically non-trivial and trivial phases, which can be characterized by Chern numbers[4]. While the experimental realizations of the electronic Chern insulators were challenging[5,6], a photonic version of the Chern insulator was proposed by Haldane[7] and experimentally observed with microwaves in a gyromagnetic photonic crystal[8,9] shortly thereafter, which opened the door to exploring topological physics in the classical-wave systems. However, this transition in photonics turns out to be difficult for acoustics because the acoustic wave doesn't respond to the external magnetic field and has no spin. Therefore, the proposals of acoustic topological insulators had to rely on the artificial gauge fields[10–12]. In addition to the complex engineering of the air flow that induces effective magnetic fields[13,14], it is also possible to generate the staggered magnetic flux in two dimensions by introducing the chiral couplings in an additional third dimension[12]. With this method, the acoustic analogue of the topological Haldane model was directly mapped and experimentally realized[15,16] later on. Since then, many ideas have been proposed along these veins and the acoustic system has become a powerful platform for exploring various topological states[17–19] in integer dimensions ranging from the quantum Hall effect[10,11], quantum spin Hall effect[20,21], valley Hall effect[22], Weyl and Dirac semimetals[15,16,23,24], high-order topological insulators[25–30], non-Abelian topological insulators[31] to aperiodic topological quasicrystals[32].

On the other hand, recent theoretical proposals show that the topological states can even go beyond the integer dimensions and exist in the fractal lattices[33–37]. Fractals appearing the same at different scales are characterized by fractal dimensions, self-similarity, and scale invariance[38,39]. They do not have a well-defined bulk like the integer-dimensional crystals and yet are able to support topological edge states. For example, it has been shown that the Floquet topological states can be implemented in a Sierpinski photonic fractal lattice of helical waveguides[37], where all the lattice sites are on the edges. The parameters adopted in this proposal are readily accessible by using direct laser writing technology[40]. These advances both in condensed matter and photonics have changed the current understanding of the bulk-edge correspondence: the topological properties do not necessarily rely on the internal bulk. Despite the increasing interest in the topological fractal insulators, the experimental realization is not straightforward and the early attempt[41] has revealed that the fractals would turn off the topological properties. Altogether, thus far, the experimental observation of the topological states in the fractals still remains elusive.

Here we present and experimentally demonstrate the squeezed Chern insulator in an acoustic fractal system. The strategy is to investigate the topological Haldane model in a fractal lattice consisting of two Sierpinski gaskets, which guarantees the same number of A and B sites with biased on-site energies. We first calculate the phase diagram based on the real-space Bott index[42] and display that the topological phase diagram is squeezed by a factor of 0.54, compared with that of the original Haldane model. The energy spectrum and eigenstates manifest the existence of topological edge states in the squeezed nontrivial regimes, whereas no such states can be found outside this nontrivial regime but within the original topological Haldane phase. Importantly, we find that contrary to the conventional topological band insulator, there exists a robust mobility gap in our fractal model protecting the topological edge states instead of the direct band gaps. Based on the theoretical model, we fabricate a synthetic acoustic system and experimentally investigate the dynamics of the topological edge states. We show that the sound waves can propagate along the outer and inner edges without penetrating into the interior of the lattice and backscattering in the presence of corners and defects. Counterintuitively, we find the topological protection can be preserved along the solely single-atom path around the defect. Our results generalize the topological Haldane model into the fractal dimensions and may open new opportunities for acoustic devices based on the topological fractal lattices.

Our starting point is a fractal lattice consisting of two Sierpinski gaskets with a fractal dimension[38] of $d_f = \frac{\ln(3)}{\ln(2)} \approx 1.585$, which is calculated from box-counting method [see Supplementary Information (SI), section 1]. Figure 1a shows the schematic of our fractal model. The red and blue dots indicate A and B sublattices, and have the same number of sites. The nearest-neighbor (NN) and next-nearest-neighbor (NNN) couplings are represented by black and grey lines. An enlarged view of the hexagonal cell in the dashed box is shown in Fig. 1b. The NNN hopping along the red or blue arrow can accumulate a phase of $\phi$. Thus, the Hamiltonian in our system can be written as

$$H = \sum_{\langle ij \rangle} t_1 c_i^\dagger c_j + \sum_{\langle\langle ij \rangle\rangle} t_2 e^{i\phi_{ij}} c_i^\dagger c_j + m(\sum_{i \in A} c_i^\dagger c_i - \sum_{i \in B} c_i^\dagger c_i) \quad (1)$$

where $c_i^\dagger$ ($c_{i,j}$) is the creation (annihilation) operator, $t_1$ and $t_2$ are NN and NNN coupling strength, $\phi_{ij}$ is the phase accumulation when hopping from site $j$ to NNN site $i$, and $m$ denotes the on-site energy difference between site A and B. The time-reversal and inversion symmetry can be broken by changing $\phi$ and $m$, respectively. In the original Haldane model, the competition between the two broken symmetries can give rise to the topological phase transition marked by the grey lines in Fig. 2. Intuitively, one may wonder whether there is a topological phase transition in our fractal system and where the phase transition is.

To verify that the topological phase transition indeed exists in our fractal model, we need to characterize the system through its topological invariant. Since the fractal lattice has no translation symmetry, we adopt an alternative invariant – Bott index[42] that is used to describe the topological properties of the disordered systems based on the real-space eigen-functions[43,44] and numerically equivalent to the Chern number in the large-scale limit[45]. The results of Bott index (see numerical details in SI, section 4) as a function of the accumulation phase $\phi$ and on-site energy difference $m$ are shown in Fig. 2. For a direct comparison, we obtain the Bott index diagram for both the honeycomb lattice (Fig. 2a) and our fractal lattice (Fig. 2b). The well-known topological transition governed by $|m| = |3\sqrt{3}t_2 \sin\phi|$ in Haldane model is marked by grey lines, as shown in both panels. These lines connect two regimes with a different Chern number $\nu$ changing from 0 to +1 or to -1. For the honeycomb lattice (inset of Fig. S2a), the topological phase diagram (Fig. 2a) calculated from Bott index shows good agreement with that calculated from Chern number. The Bott index of 0, +1, and -1 are represented by white, blue, and red colors, respectively. The transition between Bott index of $\pm 1$ and 0 overlaps with the theoretical prediction (grey lines). For the fractal lattice, the phase diagram (Fig. 2b) is significantly squeezed by 0.54 times along the vertical axis $m$, which is due to the reduction in the number of lattice sites that requires less intense inversion symmetry breaking to balance the time-reversal symmetry breaking. The phase diagrams of the different fractal generations are shown in Fig. S4 to examine the validity of the phase squeezing. These transitions are consistent with that calculated from the real-space Chern numbers (see SI, section 6). Note that despite the existence of topological phase transitions, the size of the topological bandgaps in the fractal model is relatively smaller than that in the honeycomb lattice (see SI, section 7 for details).

Having found there exists topological phase transition in the fractal model and the phase diagram is squeezed by a factor of 0.54, we calculate the eigenvalues and eigenstates at point X (non-trivial) with $\phi = \frac{\pi}{2}$, $m = 0$ and Y (trivial) with $\phi = \frac{\pi}{2}$, $m = 3.6t_2$, as shown in Fig. 3. Two points X and Y are labeled in Fig. 2b. In Fig. 3a, the eigenvalues marked by red and blue dots are the topological edge states localized on the external and internal edges. Figure 3b and 3c show the field intensities of the eigenstate 190 with the energy of -0.14 and 188 with the energy of -0.25. In Fig. 3d, the energy spectrum corresponding to point Y within the trivial regime includes no topological edge states. Figure 3e and 3f show the field intensities of the suspicious edge states with the state number of 175 with the energy of -0.88 and 176 with the energy of -0.83. However, the dynamical simulations show the states can penetrate into the interior of the fractal lattice, and these states are not topologically protected (SI, movie 4,5).

In the conventional topological insulators, the topological edge states only reside in the band gaps. However, the topological propagation in our fractal model can even survive in the presence of the trivial states with the energy close to 0 (see SI, movies 6-8). The reason is because of the lack of the well-defined bulk, there is negligible coupling between the edge states and the states localized at the perimeters of the interior voids. Therefore, there exists a robust mobility gap ranging from -0.46 to 0.46 (blue shaded region) corresponding to Bott index of 1 that protects the topological edge states instead of the direct band gaps, which makes our fractal model fundamentally distinct from the conventional topological band insulator. Note that in SI movie 9, we also show the topological transport survives as long as the perimeter of the edge encloses an area that is larger than G(3). Due to the finite-size effect, the edge enclosing a smaller area cannot support one-way transport.

To experimentally demonstrate the topological edge states in the fractal system, we design and fabricate a synthetic acoustic lattice with direct three-dimensional printing. The complex NNN hopping is introduced by adding chiral tubes in a third dimension[12]. As a result, an effective staggered magnetic flux can be included at a fixed momentum $k_z$ [$\phi = k_z d_z \in (0, \pi)$]. Figure 1c shows the enlarged hexagonal cell of our acoustic structure, which is periodic in $z$ axis and can be mapped onto the hexagonal cell in Fig. 1b. A full picture of our sample is shown in Fig. 1d and the inset shows the enlarged view of A/B meta-atoms. The detailed parameters of the structure and the measurement setup can be found in SI, sections 11 and 13. Firstly in Fig. 4a-d, we show the topological propagation of the edge states at point X with $\phi = \frac{\pi}{2}$ and $m = 0$. The schematic speakers point out the positions of the source array that can emit acoustic waves with the frequency of 11718Hz within the topological bandwidth ranging from 11109Hz to 11934Hz (SI, section 14) and a fixed momentum $k_z = \frac{\pi}{2d_z}$, which indicates the acoustic waves move not only in the x-y plane but also in the z plane (SI, section 11). After three- and five-layers propagation (Fig. 4a, b), we observe that the acoustic wave propagates along the external edge without penetration into the interior of the fractal lattice and without backscattering when encountering the sharp corner. Moving the sound source leftward makes the acoustic wave propagate through the obtuse corner and farther down along the edge, as shown in Fig. 4c-d. The red arrows point out the direction of the sound propagation. The same behavior is observed for the edge states localized at the internal edges, as shown in panels e, f. Note that because of the lossy nature of sound, the energy dissipation after propagating through each layer is about 33% and the field intensity is renormalized to the maximum value at each layer.

Furthermore, attributed to the topological protection, the acoustic wave should propagate around the defect without backscattering. We block one cavity at the position indicated by the blue arrow in Fig. 4g, h. Then we use the same method as shown in Fig. 4a, b to carry out the experiment. Clearly, we observe that the acoustic wave moves along the edge, encounters the top corner and defect, and then continues moving downward along the left edge without backscattering. We notice that around the defect that breaks the self-similarity of the system, the acoustic wave moves along a single-atom path, which pushes the robust propagation into the single-atom level. In SI, section 15, we also show in simulations that the waves can propagate around the defects located at different positions. Overall, the experimental results presented in Fig. 4 reveal that the edge states, as shown in Fig. 3a-c, are indeed topologically protected. The acoustic waves can propagate along the external and internal edges and bypass the corners and defects without backscattering, which are the key observations of these experiments.

Next, we investigate the propagation of the acoustic waves at point Y with $\phi = \frac{\pi}{2}$ and $m = 3.6t_2$, which is outside the squeezed topological regime but within the original topological phase enclosed by grey lines. The on-site energy difference $m$ between A and B sublattices is introduced by simply tuning the height of the sound cavities (SI, section 16). After five-layer propagation, as shown in Fig. 5a, b, we can see that the acoustic waves penetrate into the interior of the sample and scatter at the corners, as the source moves from the upper right edge to the lower left edge. The operating frequency is 10506Hz corresponding to the energy of the suspicious edge states. The trapped behavior is also observed for sound propagating along the perimeter of the center void (Fig. 5c), which is in sharp contrast to the result shown in Fig. 4f. Combined with the results at point X as shown in Fig. 4, we have experimentally verified that introducing the fractal lattice into the topological Haldane model retains the topological phase transitions, whereas the topological phase diagram is compressed.

In summary, we proposed and experimentally demonstrated an example of acoustic topological fractal insulators, which supports the topological edge states protected by a mobility gap that is distinct from the common topological band insulator. We pinpointed that the topological phase diagram is compressed by about 0.54 times, compared with that of the original Haldane model. In the experiments, we observed the robust propagation along the outer and inner edges of our fractal lattice. Surprisingly, the topological protection is preserved in the single-atom path around the defect localized on the edges of the fractal lattice. Our findings open the door to exploring rich topological physics ranging from the first-order to higher-order topological insulators in the classical-wave fractal systems[17,46,47]. Meanwhile, many interesting questions are raised, for example, the possibility of pursuing the topological states in the random fractals[38] that are different from the current deterministic fractals such as Sierpinski gasket and carpet, the interplay between topology and non-Hermiticity[48–50] in the fractals, and potential applications such as topological fractal insulator sensors[51]. The realization of a topological fractal insulator in our accessible acoustic system will help us to address these questions.

**Supplementary Information** is available in the online version of the paper.

**Acknowledgements**
We acknowledge the support of the National Natural Science Foundation of China (Grants No. 12174339) and Fundamental Research Funds for the Central Universities.

**Author contributions**
All authors contributed substantially to this work.

**Competing interests**
The authors declare no competing interests.

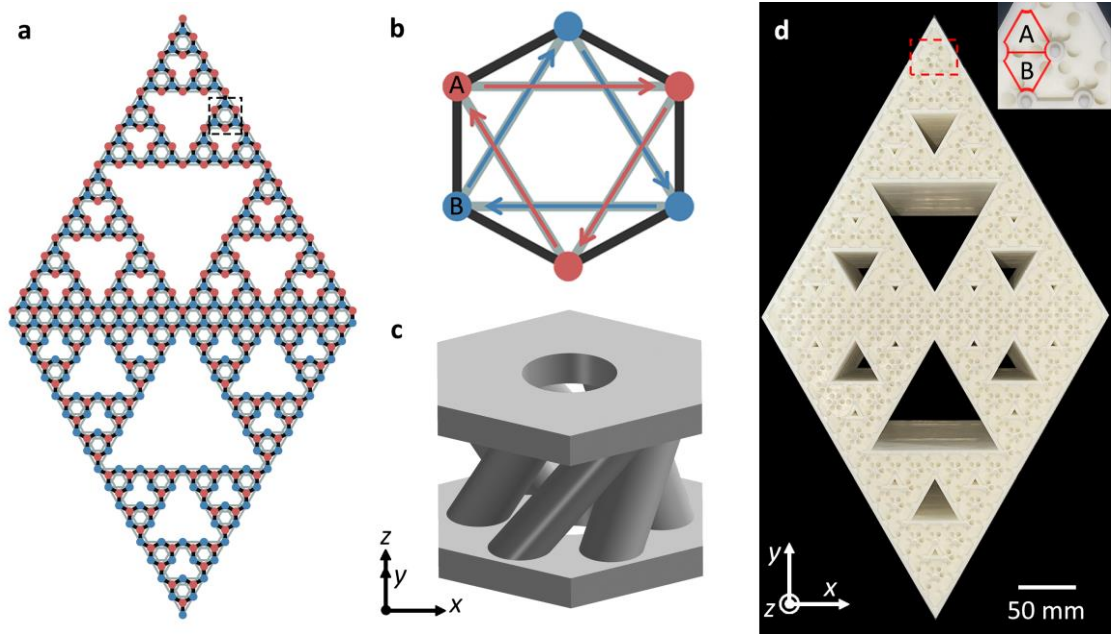

**Fig. 1 | The schematic and sample photo of our fractal model. a**, The schematic of the fractal lattice that is composed of two Sierpinski gaskets. Red and blue dots indicate A and B sublattices. The black (grey) lines represent NN (NNN) couplings. **b**, An enlarged view of a hexagonal cell corresponding to the dashed box in **a**. The NNN hopping along the red or blue arrow can accumulate a phase of $\phi$, and vice versa. **c,** An acoustic structure of the hexagonal cell corresponding to panel **b**. The complex NNN hopping is introduced by adding chiral tubes with fixed $k_z$. **d**, The photo of our acoustic system. The inset shows an enlarged view of A and B meta-atoms. The on-site energy difference between A and B meta-atoms can be tuned by changing the height of the cavities.

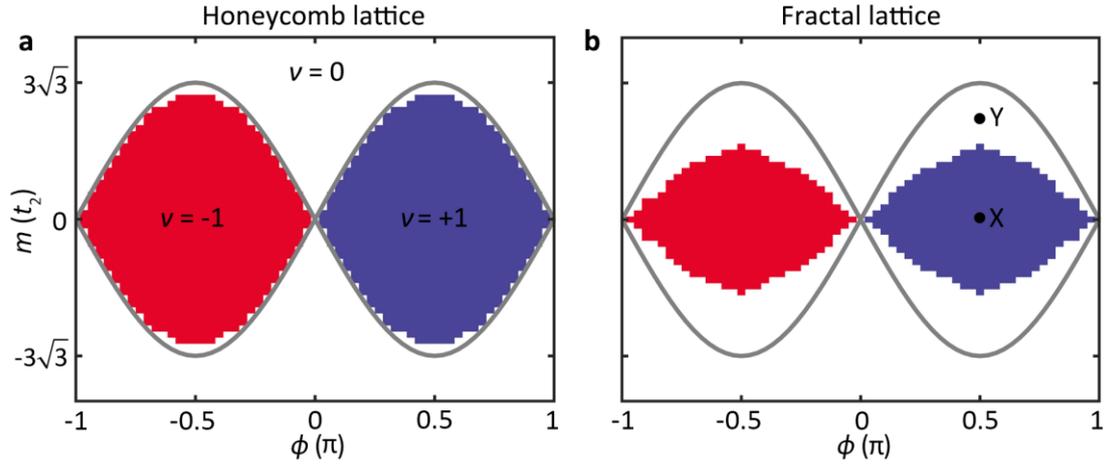

**Fig. 2 | Topological phase diagram of the rhombic honeycomb and fractal lattices. a**, Topological phase diagram of the rhombic honeycomb lattice. In the original Haldane model, topological phase transitions obey $|m| = |3\sqrt{3}t_2 \sin\phi|$, which is marked by the grey lines. The Bott index of 0, +1, and -1 are represented by white, blue, and red colors, respectively. The shaded regions calculated from Bott index are consistent with the regions having nonzero Chern numbers. **b**, The phase diagram of the fractal lattice is significantly squeezed by 0.54 times along the vertical axis $m$. The coupling parameters for theoretical calculations are: $t_1 = 1$, $t_2 = 0.2$.

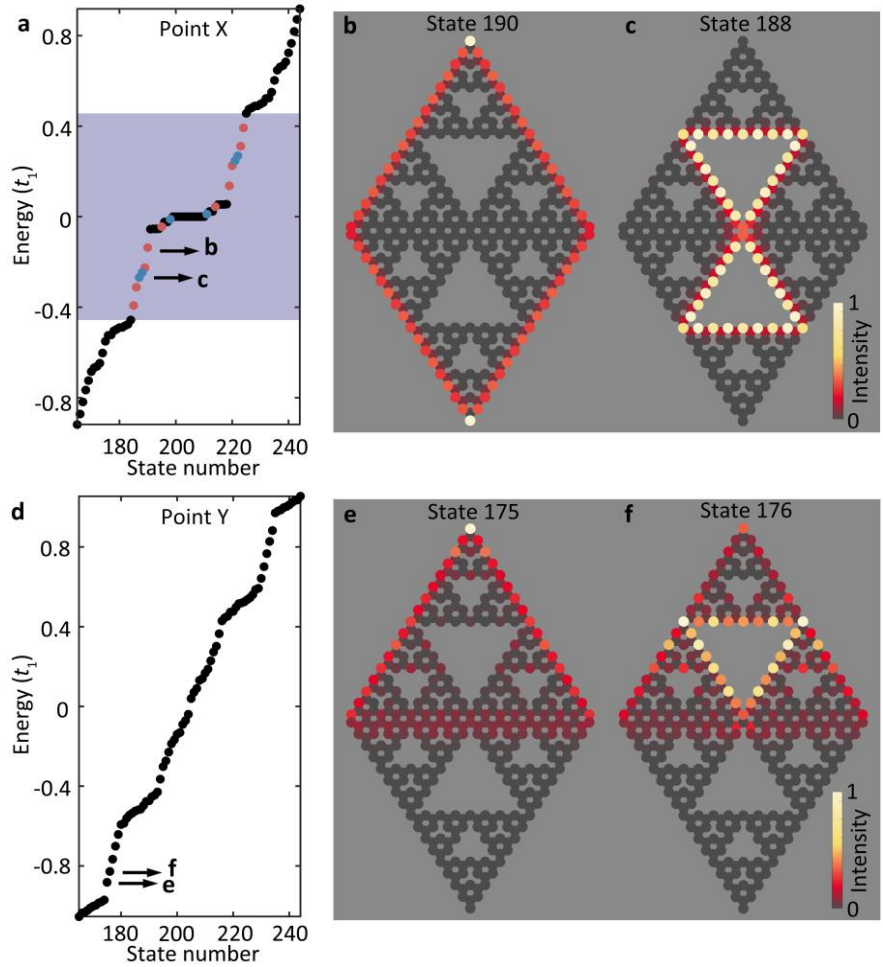

**Fig. 3 | The energy spectra and edge states of the fractal lattice. a**, **d**, The energy spectra for the topological nontrivial (a) and trivial (d) phases, which correspond to point X and Y as labeled in Fig. 2b, respectively. The red and blue dots indicate the external and internal edge states. The blue shaded region indicates the topological bandwidth, where the topological transmission can be supported. **b**, **c**, The field intensities of the edge states localized at the external and internal edges, which correspond to the state number 190 with energy of -0.14 and 188 with energy of -0.25 (SI, movie 1,2). **e**, **f**, The field intensities of the eigen state 175 with energy of -0.88 and 176 with energy of -0.83. These two states are the suspicious edge states. However, we confirm by simulations that the waves will backscatter and penetrate into the interior of the lattice. The parameters for the calculations are the same as in Fig. 2

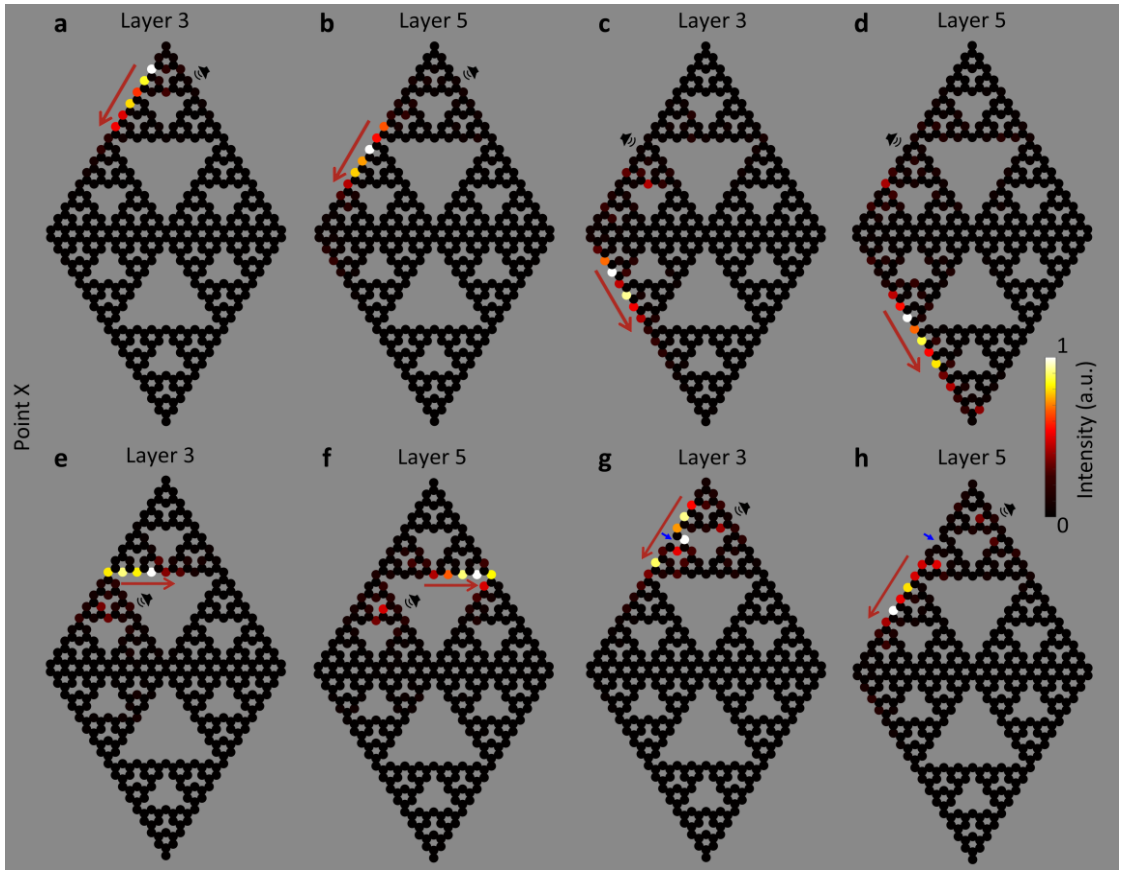

**Figure 4 | Experimental observation of the topologically protected edge states in the acoustic fractal lattices.** **a**, **b**, Sound intensity distribution of the external topological edge states at propagation layer 3 and 5. **c**, **d**, Moving the sound source leftward makes the acoustic wave propagate through the obtuse corner and farther down along the edge. **e**, **f**, Intensity distribution of the internal topological edge states at propagation layer 3 and 5. **g-h**, Intensity distribution of the edge states at propagation layer 3 and 5 in the presence of a defect indicated by blue arrows. We can see from the experimental results that the acoustic waves can propagate along the external and internal edges and bypass the corners and defects without backscattering. Note that the topological systems in this figure correspond to point X labeled in Fig. 2b. The schematic speakers indicate the positions of the source array that emits acoustic wave with the frequency of 11718Hz and a fixed momentum $k_z = \frac{\pi}{2d_z}$. The red arrows point out the direction of the sound propagation.

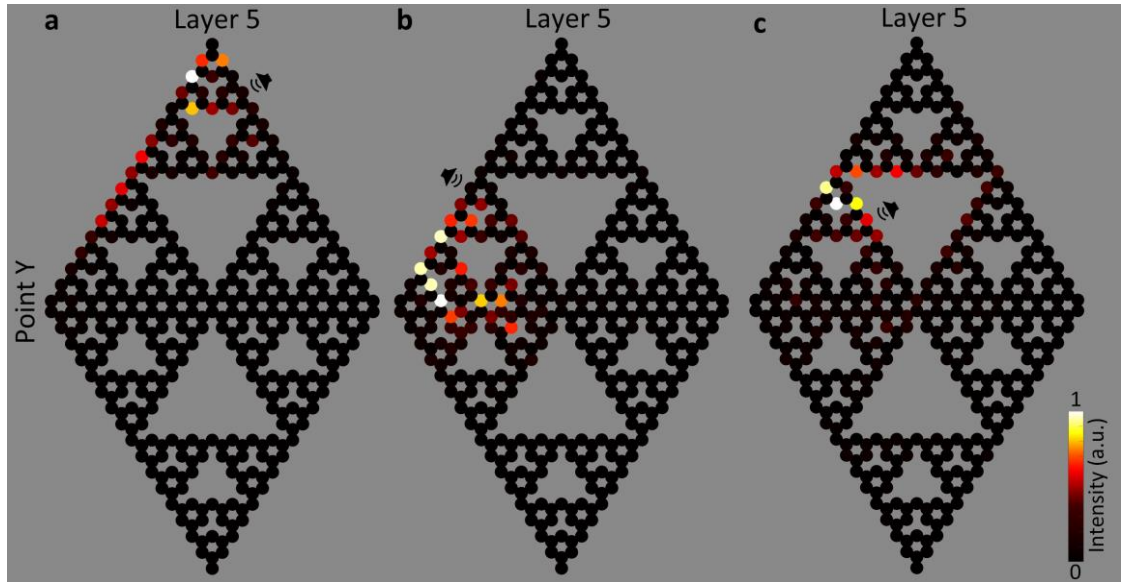

**Figure 5 | Experiments showing the trapped sound in the trivial phase. a**, **b**, Sound intensity distribution of the acoustic waves launched at the external edges at propagation layer 5 with two different source positions. **c**, Intensity distribution of acoustic wave launched at the internal edge at propagation layer 5. We can see that the acoustic waves cannot propagate along the edges and bypass the corners. Note that the trivial system in this experiment corresponds to point Y as labeled in Fig. 2b. The schematic speakers indicate the positions of the source array that emits acoustic waves with the frequency of 10506Hz for **a**, **b** and **c**. The output source array has a fixed momentum $k_z = \frac{\pi}{2d_z}$.